# Deep-UV to mid-IR supercontinuum generation driven by mid-IR ultrashort pulses in a gas-filled fiber

ABUBAKAR I. ADAMU,[1,\*\*] MD. SELIM HABIB,[2,\*\*] CHRISTIAN R. PETERSEN,[1,\*\*] J. ENRIQUE ANTONIO-LOPEZ,[2] BINBIN ZHOU,[1] AXEL SCHÜLZGEN,[2] RODRIGO AMEZCUA-CORREA,[2] OLE BANG,[1] AND CHRISTOS MARKOS[1\*]

[1]*DTU Fotonik, Technical University of Denmark, Kgs. Lyngby, DK-2800, Denmark*
[2]*CREOL, The College of Optics and Photonics, University of Central Florida, Orlando, FL-32816, USA*
*\*\*equal contribution*
*[\*]chmar@fotonik.dtu.dk*

**Abstract:** Supercontinuum (SC) generation based on ultrashort pulse compression constitutes one of the most promising technologies towards an ultra-wide bandwidth, high-brightness and spatially coherent light sources for applications such as spectroscopy and microscopy. Here, multi-octave SC generation in a gas-filled hollow-core antiresonant fiber (HC-ARF) is reported spanning from 200 nm in the deep ultraviolet (DUV) to 4000 nm in the mid-infrared (mid-IR). A measured average output power of 5 mW was obtained by pumping at the center wavelength of the first anti-resonance transmission window (2460 nm) with ~100 fs pulses and an injected pulse energy of ~7-8 µJ. The mechanism behind the extreme spectral broadening relies upon intense soliton-plasma nonlinear dynamics which leads to efficient soliton self-compression and phase-matched dispersive wave (DW) emission in the DUV region. The strongest DW is observed at 275 nm having an estimated pulse energy of 1.42 µJ, corresponding to 28.4 % of the total output energy. Furthermore, the effect of changing the pump pulse energy and gas pressure on the nonlinear dynamics and their direct impact on SC generation was investigated. The current work paves a new way towards novel investigations of gas-based ultrafast nonlinear optics in the emerging mid-IR spectral regime.

## 1. Introduction

The ultraviolet (UV) spectral range is of great scientific and technical interest, mainly because a majority of molecules have strong electronic bands in this region [1], while the mid-IR on the other hand is directly associated with the strong fundamental vibrational resonances of polarized (polyatomic) molecules that have distinctive spectral fingerprints [2,3]. Triggered by a tremendous number of applications such as semiconductor metrology and inspection [4], pump-probe spectroscopy [5], pollution monitoring [6], optical coherence tomography and imaging [7,8], fiber-based laser sources capable of covering the electromagnetic spectrum from DUV up to mid-IR, have attracted the scientific attention of several research groups around the globe. One of the most promising routes towards development of ultra-broad bandwidth sources is SC generation. The main nonlinear effects and physics behind SC generation have been well-studied in the past few decades, rendering SC generation to be a well-established and matured technology [9,10]. However, most reports on solid-core fibers in the literature are based on silica photonic crystal fibers (PCF) in which, the microstructured cladding of the fiber allows tailoring of the group velocity dispersion (GVD), which is an important property that greatly influences the nonlinear effects [10].

Despite silica solid-core PCF-based SC sources now being commercially available, they can only reliably operate in a limited transmission range around 350-2300 nm, restricted by the transparency of the fiber material [11]. In 2015, Jiang *et al.* reported for the first time the fabrication of a fluoride (ZBLAN) glass-based solid-core PCF with high air-filling fraction and they demonstrated a broad SC spanning more than three octaves in the spectral range 200-2500 nm [12]. However, fabrication of ZBLAN-based PCF remains still a very challenging task due to the very steep viscosity–temperature profile of the glass even for the fabrication of conventional step-index fibers. In order to overcome the limitations imposed by the fiber materials, a relatively new research subfield has been created in nonlinear fiber optics using gas-filled hollow-core PCFs instead of solid-core fibers [13–15].

Gas-based nonlinear optics in fibers and capillaries is certainly not a new field. Ippen for example has demonstrated backward stimulated Raman scattering in a glass capillary fiber in 1972 [16]. Since then, gas-filled large bore capillaries have been a standard approach for nonlinear applications such as pulse compression [17] and high-harmonic generation [18]. With the invention of hollow-core PCF, new opportunities appeared because it became possible to have a low-loss waveguide with core diameters of tens of micrometers, thus allowing the initiation of nonlinear effects through light-gas interaction at much lower pulse energies compared to capillaries [13,19]. Hollow-core PCFs and in particularly antiresonant hollow-core fibers (AR-HCF) are divided into two main categories based on their geometrical structure: Kagome and negative-curvature [20]. The latter is a simplified form of the former which has lately received a lot of scientific attention from the fiber-optic community because it offers flatter transmission spectrum, guidance in the mid-IR region and reduced complexity from a fabrication point of view [15,20]. Nevertheless, both fiber structures have been extensively reported in the literature for gas-based experiments. The main reasons are because they offer a weak anomalous dispersion which compensates the normal dispersion of the filling gas but more importantly they can tolerate extremely high terawatt levels of peak power due to low modal overlap with silica. Many impressive results, such as tunable broadband deep- and vacuum UV light generation by tuning the gas pressure inside the fiber [13,14,21–24], four-wave mixing [25], and Raman-induced multi-octave SC generation have been demonstrated [26–28]. Most of the experimental papers so far report on pumping a gas-filled AR-HCF in the visible and near-infrared regime close to the zero-dispersion wavelength (ZDW). Recently, Cassataro *et al.* reported that pumping a Krypton-filled AR-HCF under 18 bar pressure at 1.7 μm using an optical parametric amplifier is enough to generate SC from 270 nm up to 3.1 μm with a total output energy of 4 μJ [29]. However, to the best of our knowledge there has been no experimental research carried out with pumping a gas-filled AR-HCF with ultrashort pulses in the mid-IR spectral region due to the lower photon energy available for plasma initiation [30,31].

This work reports on the first experimental demonstration of multi-octave SC generation in a gas-filled AR-HCF pumped in the mid-IR region at 2460 nm using a tunable optical parametric amplifier (OPA) system. By coupling 100 fs, 20 μJ pulses into a specially designed argon (Ar) filled AR-HCF under 30 bar pressure, soliton self-compression dynamics enabled broadening from 200 nm to 4000 nm. Furthermore, it was experimentally demonstrated how the pulse energy and the pressure have a crucial role in the mid-IR spectrum broadening and emission of DWs in the DUV.

## 2. Experimental and theoretical background

### 2.1 Experimental configuration and HC-ARF

The HC-ARF used in this experiment was specially designed for the mid-IR region. The fiber consists of a hollow core surrounded by seven non-touching silica capillaries with wall thickness of ~640 nm forming a core with diameter of ~44 μm, as shown in the scanning electron microscopy (SEM) image of Fig. 1 (a). The calculated fundamental mode profile at 2.46 μm (pump wavelength) and at the first resonance (1.38 μm) is also shown in Fig. 1 (a) The 30 cm long HC-ARF was suspended between the two gas cells, equipped with 5mm $CaF_2$ windows for input fiber coupling and collimation at the output.

Figure 1 (b) shows the experimental setup used in this work. A Ti:sapphire amplifier was used to pump an OPA with a few milijoule pulse energy. Linearly polarized, 100 fs, mid-IR pump pulses with central wavelength at 2460 nm, with 1 kHz repetition rate were generated from the OPA. The mid-IR pump power was controlled by neutral density (ND) filters and by rotating a nanoparticle linear film polarizer (P1). The light is free-space-focused into the fiber with an aspheric lens with 50 mm focal length and a similar aspheric lens collimates the light at the output of the fiber. Both windows and lenses were uncoated to obtain >90% transmission from 200-5000 nm. The beam was directed to a CCD-based detector and a flip mirror is used to direct the beam to the IR Spectrometer. The fiber output power was measured using a thermal power meter, and the beam near-field profile was imaged using a visible and near-IR CCD camera. The output spectrum was measured from 183-1100 nm using a fiber-coupled UV-VIS CCD array spectrometer (Oceanoptics HR2000+), and spectrum from 1000-5000 nm was measured using a scanning spectrometer equipped with a mercury-cadmium-telluride (MCT) detector and box-car integrator. The scanning spectrometer included an automatic long-pass filter-wheel to eliminate higher-order diffraction, and the spectral response of the entire system was calibrated using a 1273 K blackbody source.

The gas cells were connected with plastic tube for gas entry, and a valve for purging and gas evacuation. The fiber was mounted and fixed to a v-groove inside the gas-cell to ensure stability at high pressure, and the two gas-cells were mounted on micro-translation stages and moved in parallel to avoid fiber bending while allowing for flexible input coupling. The insertion loss of the fiber and $CaF_2$ optics was 4.2 dB (38 % transmission) for a central pump wavelength of 2460 nm. The output power was measured for every measurement, and coupling was optimized after each change in pressure and power to maintain efficient coupling to the fundamental mode and to prevent damage of the facet. The gas used for the experiments is compressed Ar (99.998% purity), and a pressure of up to 30 bar was applied into the fiber from both gas-cells to maintain uniform pressure in the system. The pressure was varied using a manual high-pressure regulator. The measurements were stopped a few seconds after every change of pressure to ensure uniform distribution of pressure in the tubes, gas-cells and the fiber.

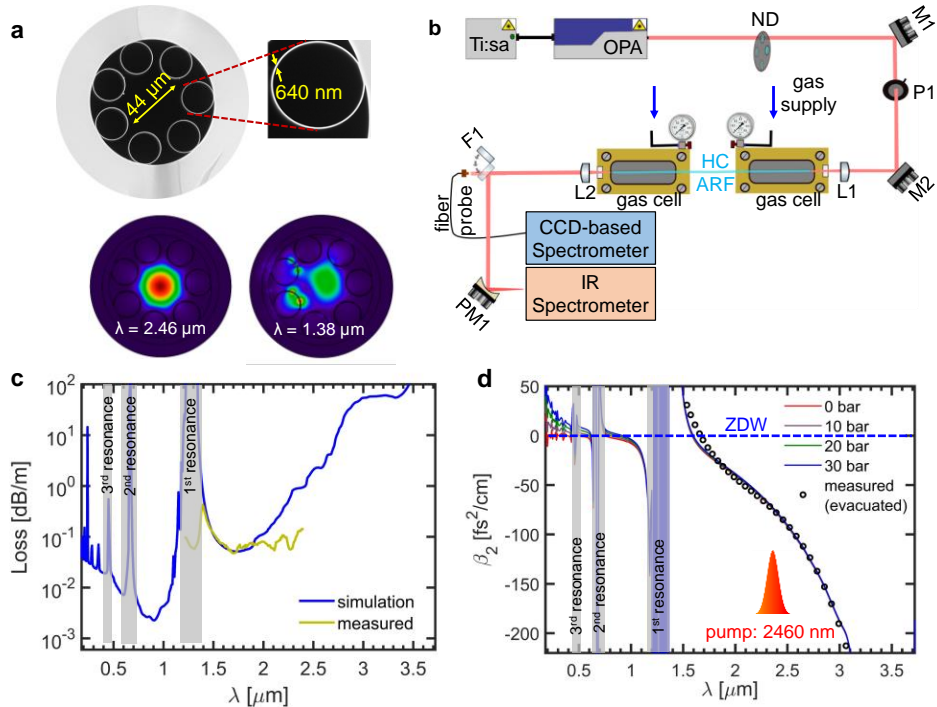

Fig.1. (a) SEM of the HC-ARF (top panel); and mode-field profile at 2460 nm (pump wavelength) and 1380 nm (1$^{st}$ resonance) (bottom panel). The fiber has a core diameter ~44 μm, an average capillary diameter 25 μm, and average wall thickness of 640 ± 50 nm. (b) Experimental setup for the gas-filled HC-ARF nonlinear experiment. The components of the set up are: neutral density filter (ND), protected silver mirrors (M1-3), linear film polarizer (P1), CaF$_2$ plano-convex lenses (L1-2), , flip mirror (F1), and a gold coated parabolic mirror (PM1), HC-30 cm of ARF,. (c) Calculated losses (blue) for the HC-ARF including contributions from surface scattering, mode confinement, and silica material loss and measured propagation loss (yellow). (d) Calculated GVD for the HC-ARF at various levels of Ar gas pressure (solid lines). The measured dispersion of the evacuated fiber in the mid-IR is shown for comparison (circles).

Figure 1 (c) shows the calculated total propagation loss of the HC-ARF used in the experiments based on a Finite Element Method (FEM) software with contributions from the material (silica), mode confinement, and surface scattering loss (SSL). The yellow line is the measured propagation loss, which relatively agrees with the calculated one. As it can be seen from Fig. 1(c), the HC-ARF transmits light over a broad range of wavelengths with the first antiresonance window to be from ~1400 nm to ~4000 nm (indicated as 1$^{st}$ resonance in Fig. 1(c). Higher order anti-resonant windows allow for light to propagate even in the DUV range close to the bandgap of fused silica. The refractive index used in the calculations was based on the Sellmeier equation found in [32], which is valid from 210 nm up to 3.7 μm. However, it is important to mention that careful consideration of the silica refractive index is required for wavelengths less than 210 nm due to strong electronic bandgap absorption [33]. To this end, the measured refractive index data of silica from [33] was used to calculate the GVD and confinement loss for wavelengths less than 210 nm. Figure 1(d) shows the calculated GVD of the fiber from 200 nm to 3700 nm for different gas pressures from 0 to 30 bars. When the pressure of the gas in the fiber increases, the ZDW shifts towards longer wavelengths, and the nonlinear refractive index n$_2$ increases as it has been extensively described elsewhere [13–15]. In order to verify the modelling, the GVD was also measured for the evacuated HC-ARF in the spectral range from 1.5 μm to 4 μm using a mid-IR SC source (mid-IR superK compact, NKT Photonics A/S). The calculated and measured dispersion profiles are in very good agreement thus confirming the accuracy of the calculations.

## 2.2 Modeling and theory

The single-mode optical pulse propagation in gas-filled HC-AR fiber was studied using a unidirectional field equation which also accounts for free-electron effects expressed as [31,34,35]

$$\partial_z E(z,\omega) = i(\beta(\omega) - \frac{\omega}{v_p} + i\frac{\alpha}{2})E(z,\omega) + i\frac{\omega^2}{2c^2\beta(\omega)}F(P_{NL}(z,t)), \qquad (1)$$

where $z$ is the propagation distance along the fiber, $t$ is the time in the reference frame moving with the pump group velocity $v_p$, $E(z,\omega)$ is the electric field in the frequency domain, $\omega$ is the angular frequency, $\alpha(\omega)$ is the linear propagation loss of the fiber, $c$ is speed of light in vacuum speed, $\beta(\omega)$ is the propagation constant, and $F$ represents the Fourier transform. $P_{NL}(z,t)$ is the nonlinear polarization calculated by [34,35] $P_{NL}(z,t) = \varepsilon_0 \chi^{(3)} E(z,t) + P_{ion}(z,t)$, The first term is the Kerr effect, where $\varepsilon_0$ is vacuum permittivity, and $\chi^{(3)}$ is the third-order nonlinear susceptibility. The second term describes the nonlinear polarization due to the ionization effect calculated as [35–38]

$$P_{ion}(z,t) = \int_{-\infty}^{t} \frac{\partial \rho}{\partial t'} \frac{I_p}{E(z,t')} dt' + \frac{e^2}{m_e} \int_{-\infty}^{t} \int_{-\infty}^{t'} \rho(z,t'')E(z,t'')dt''dt', \qquad (2)$$

where $\rho$ is density of free electrons, $m_e$ and $e$ are the mass and charge of an electron, and $I_p$ is gas ionization energy. In our calculation, nonlinear refractive index ($n_2$) of Ar was assumed wavelength independent [31,39] and also neglect the Raman contribution due to the very low light-glass overlap (<<1%) in HC-AR fiber [31,40,41]. For gas-filled fibers, the optical pulse intensity in the range of 100 TW/cm$^2$ corresponds to Keldysh parameter, $p_k \leq 1$ [42,43]. In our gas-filled system, the peak intensity reaches 190 TW/cm$^2$ confirming $p_k <<1$ in which tunnel ionization dominates over multiphoton ionization [31]. Experimental results also confirm that tunneling model provides excellent agreement even with $p_k \simeq 1$ [44,45].Therefore, to calculate the free electron density, quasi-static tunneling ionization was chosen based on the Ammosov, Delone, and Krainov (ADK) model which is described in [42].

The evolution of spectral and temporal profile in a 44 μm core HC-AR fiber filled with 30 bar Ar with 8 μJ energy and pumped in the anomalous dispersion regime at 2460 nm with 100 fs Gaussian pulse are shown in Fig. 2(c) and Fig. 2(e), respectively. After propagating 7.1 cm, the pulse experiences a strong soliton self-compression down to ~1.6 fs (less than a single cycle) with peak intensity ~190 TW/cm$^2$ which is due to the combined action of SPM, anomalous dispersion, and optical shock effects [31,35] (see Fig. 2(e)). At the maximum temporal compression point, an efficient dispersive wave (DW) is emitted at 278 nm which calculated efficiency of >15% (see Fig. 2(d)). To confirm the DW at 278 nm, the propagation constant mismatch Δβ was calculated, between the soliton and dispersive waves which is shown in Fig. 2(a) and calculated by [35,36]

$$\Delta\beta(\omega) \approx \beta(\omega) - (\beta_0 + (\omega - \omega_0)\beta_1 + \gamma N P_0 \, \omega/\omega_0 - \frac{\omega_0 \rho \omega_0}{2 n_0 c \rho_{cr} \omega}), \quad (3)$$

where, $\beta_0$ is the propagation constant and $\beta_1 = 1/v_g$ is the inverse group velocity at $\omega_0$, $P_0$ is the pump peak power, $N$ is the soliton order, $\gamma = n_2/cA_{\text{eff}}$ is the fiber nonlinear coefficient, $A_{\text{eff}}$ is the effective mode area of the fiber, $\rho$ is the plasma density and $\rho_0$ the critical plasma density at which plasma becomes opaque [35,36]. In the spectral profile, a narrow band emission peak was found at the resonance wavelength of ~1360 nm despite the high loss at this wavelength. This feature at the vicinity of the anti-crossing is a consequence of the phase-matching curve crossing $\Delta\beta = 0$ resulting in four-wave mixing and the DW emission, as recently reported in Ref. [46,47]. It can be seen from Fig. 2(c) that a blue shifted soliton is found at the maximum temporal compression point because such high peak intensity is enough to ionize the gas and form a plasma (see Fig. 2(c)). Due to the plasma formation, the spectrum broadens mainly to the blue side and a multi-octave supercontinuum is found covering 200 nm to above 4000 nm. At the soliton self-compression stage, the average nonlinear index change becomes negative which in turn provides a positive phase shift and a shift towards blue side of the spectrum whereas the index change remains positive before and after the self-compression [48]. After the self-compression, the peak intensity falls below the threshold value sufficient to ionize the gas for additional ionization [42]. It is important to emphasize that these simulations assume a spatially invariant fundamental mode, while it is known that such ultra-high peak power can lead to beam filamentation through a combination of Kerr self-focusing and plasma defocusing. As a result the true spatio-temporal dynamics present a highly complex problem that requires more elaborate models. The implementation of the numerical model presented here is described in ref. [31].

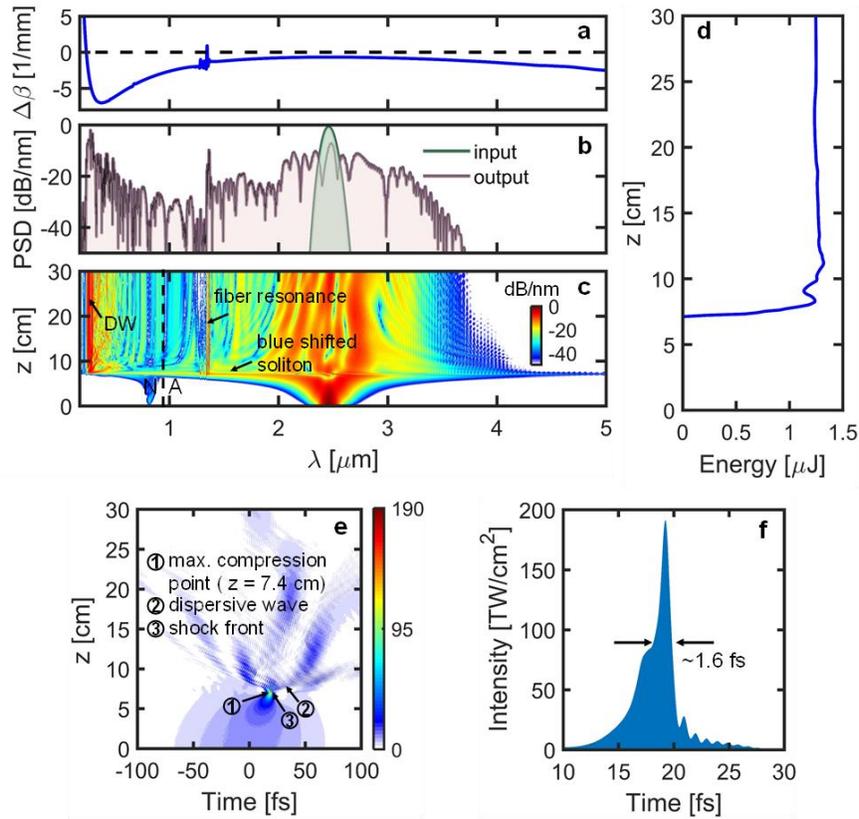

Fig. 2. Numerically simulated (a) propagation constant mismatch ($\Delta\beta_{mismatch}$) between solitons and dispersive waves, (b) normalized power spectral density (PSD) at z = 0 and z = 30 cm, (c) spectral evolution (normalized to the peak intensity), (d) energy of the DW, (e) temporal evolution in TW/cm$^2$, and (f) intensity at the maximum compression point at z = 7.1 cm for a ~44 μm core HC-AR fiber under 30 bar filling with Ar, pulse energy 8 μJ, pulse duration 100 fs pumping in the anomalous dispersion regime at 2460 nm. N: normal dispersion regime, and A: anomalous dispersion regime. Fiber dispersion was calculated using finite-element modeling.

## 3. Results and discussion

Figure 3 (a) shows the comparison of the measured and calculated SC spectra spanning from the 200 nm up to 4 μm. Before the start of the experiments, the fiber was purged several times with high purity Ar (AGA A/S) to remove any remaining atmospheric air and other impurities. The fiber length was 30 cm to ensure sufficient light-gas interaction, in order to initiate the described nonlinear effects. At the maximum pump power of 20 mW, corresponding to a pulse energy of 20 μJ, a total of 5 mW average power was measured out of the fiber after the collimation lens. By taking into account the loss of the CaF$_2$ optics, which has 94 % transmission at 2460 nm and ~92 % transmission on average from 200-4000 nm (source: Thorlabs) and including around 2 dB/m propagation loss at the pump wavelength, the 5 μJ output energy correspond to an estimated injected pulse energy of around 8 μJ. This estimate is corroborated by the qualitative agreement between experiments and simulations for an injected pulse energy of 8 μJ.

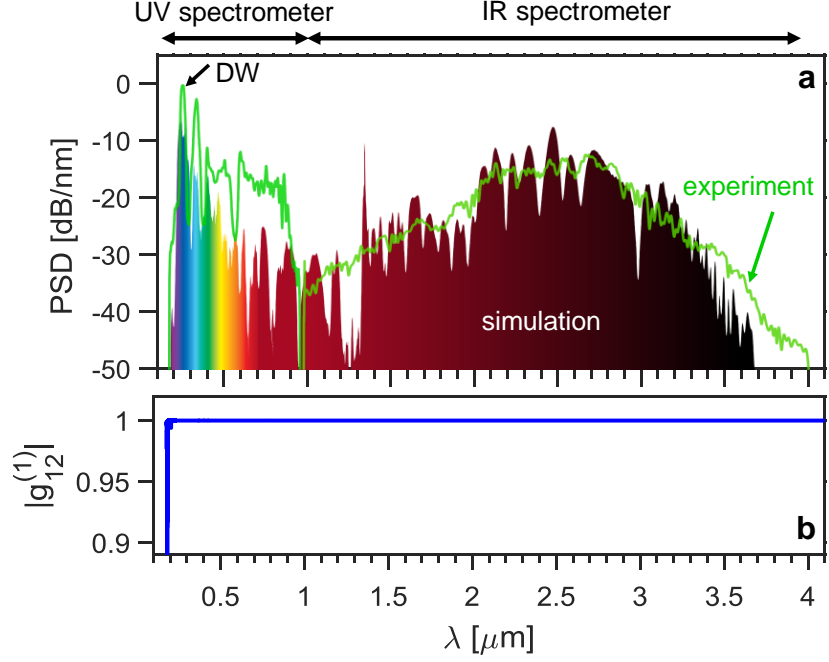

Fig. 3. (a) Simulation and measurement of the SC spectrum pumped at 2460 nm (anomalous dispersion region) with pulse duration 100 fs pumping (b) Complex degree of coherence of the generated SC.

The complex degree of first-order coherence of the supercontinuum is shown in Fig. 4(b) which is calculated using [24,49]

$$\left|g_{12}^{(1)}\right| = \left|\frac{\left\langle A_1^*(\omega)A_2(\omega)\right\rangle}{\left\langle \left|A_1(\omega)\right|^2\right\rangle}\right|, \tag{4}$$

Where the angle brackets represent an ensemble average over the independent simulations. The value of $\left|g_{12}^{(1)}\right|$ indicates quality of the coherence of the supercontinuum and the coherence is perfect if $\left|g_{12}^{(1)}\right| = 1$, meaning that the electric fields have perfectly equal phase from different laser shots whereas $\left|g_{12}^{(1)}\right| = 0$ indicates the random phase fluctuation from shot-to-shot. Figure 4(b) depicts that the first-order coherence of the full generated spectrum is fully coherent $\left|g_{12}^{(1)}\right| \approx 1$.

The main limitation of the spectral extension towards the mid-IR is the increasing propagation losses, reaching 100 dB/m at 3500 nm from which the silica material loss is the dominant loss mechanism. This was also the conclusion from the numerical simulations, as seen in Fig. 2(b). Figure 4(a) shows the mid-IR spectral broadening with increasing pump power at a maximum pressure of 30 bar where the nonlinearity is high. At lower pump power (1-3 mW output power) the limited broadening observed can be attributed to SPM. In this low-power regime the mid-IR broadening is only weakly dependent on pressure. However, as seen in Fig. 4(b) the visible part of the spectrum was highly influenced by changes in the pressure. This is to be expected from the significant change in the GVD in the visible compared to the mid-IR. At higher power, the pressure begins to have a more pronounced effect on the mid-IR and UV-visible regime. These dynamics are shown for 5 mW output power in Fig. 4 (c) using a base 2 logarithmic wavelength axis for better visibility of the UV region. Above 15 bar the long-wavelength edge starts to increase together with the energy of the DWs around 280 nm, while simultaneously diminishing the energy of the DWs around 240 nm. The increasing mid-IR broadening and 280 nm DW generation is attributed to the increasing nonlinearity, while the diminishing 240 nm DW generation is believed to be caused by the red-shifting of the ZDW.

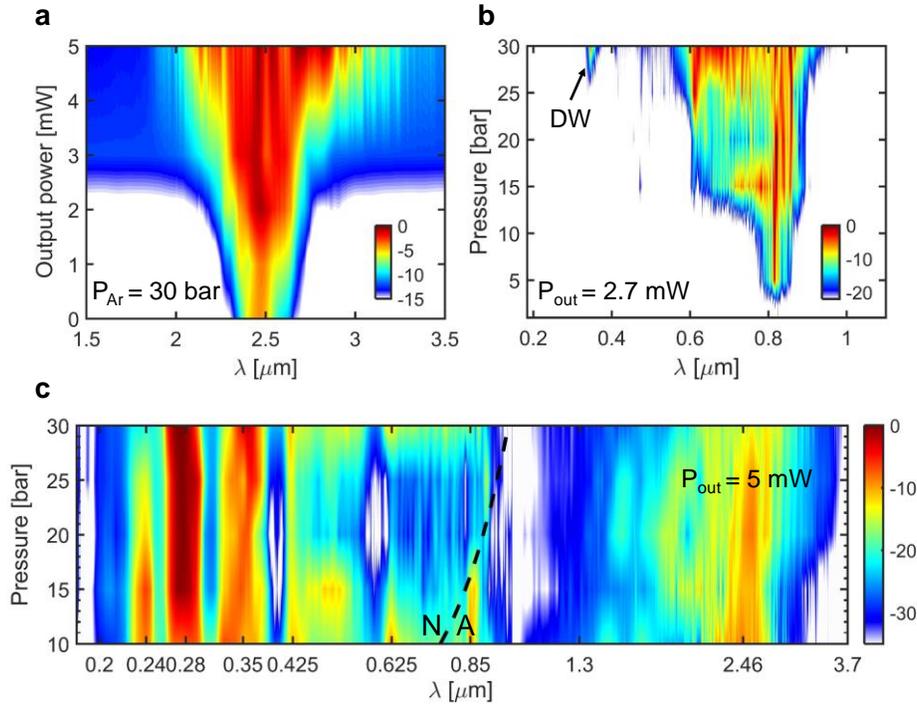

Fig. 4. (a) Measured spectral evolution and DW formation in the near/mid-IR range indicating the spectrum broadening as a function of measured output power for a fixed pressure of 30 bar. (b) Spectral broadening and DW emission as a function of pressure in the DUV/visible. (c) Pressure dependent evolution of the spectrum at a fixed low power of 5 mW over the full spectrum from 10 up to 30 bars with a step of 5 bars.

In conclusion, multi-octave SC generation was demonstrated in a gas-filled HC-ARF pumped in the mid-IR region. The SC spectrum spans from 200 nm up to 4.0 μm when 8 μJ, 100 fs pulses are injected into the fiber filled with Ar at a pressure of 30 bar. A total measured average output power of 5 mW was obtained with a resonant DW emission at 275 nm having an estimated pulse energy of 1.42 μJ corresponding to 28.4% of the total output energy. Finally, it was experimentally demonstrated how the pump energy and pressure increases the nonlinearity resulting in increased mid-IR spectral broadening and efficient DW emission in the DUV range. The current work constitutes an efficient route towards ultrafast source for spectroscopy both in the mid-IR molecular fingerprinting and in the DUV spectral region.

**Funding**

Innovation Fund Denmark (6150-00030B); Det Frie Forskningsråd (DFF) (4184-00359B); Army Research Office (ARO) (W911NF-17-1-0501 and W911NF-12-1-0450); Air Force Office of Scientific Research (AFOSR) FA9550-15-10041).

**Acknowledgements**

The authors would like to thank Morten Bache for fruitful discussions and Manoj Dasa for assistance with Figure 1 (b).